\documentclass{article}
\usepackage{spconf,amsmath,graphicx}
\usepackage{mathrsfs,hyperref}
\usepackage{stackengine}
\usepackage{algorithm}
\usepackage{bm}
\usepackage{siunitx}
\usepackage{subcaption}
\usepackage{algpseudocode}\usepackage{xspace}
\usepackage[export]{adjustbox}
\usepackage{xcolor}
\usepackage{makecell}

\makeatletter
\DeclareRobustCommand\onedot{\futurelet\@let@token\@onedot}
\def\@onedot{\ifx\@let@token.\else.\null\fi\xspace}

\def\eg{\emph{e.g}\onedot} 
\def\ie{\emph{i.e}\onedot}

\def\etal{\emph{et al}\onedot}
\makeatother

\algnewcommand{\LineComment}[1]{\State \(\triangleright\) #1}

\usepackage[capitalize]{cleveref}
\crefname{section}{Sec.}{Secs.}
\crefname{section}{Section}{Sections}
\crefname{table}{Table}{Tables}
\crefname{table}{Tab.}{Tabs.}
\crefname{algorithm}{Alg.}{Algs.}

\newcommand{\insertwithsubimagenew}[3][200 120 190 180]
{\stackinset{l}{-0.1cm}{t}{-0.1cm}
  {\scalebox{1}{\includegraphics[trim=#1,clip,width=1.3cm]{#3}}}
  {\includegraphics[width=#2]{#3}}}

\title{A Modular and Robust Physics-Based Approach for Lensless Image Reconstruction}
\name{Yohann Perron$^*$, Eric Bezzam$^*$, Martin Vetterli\thanks{
$^*$These authors contributed equally to this work.\\
This project was supported by the Open Research Data Program of the ETH Board, and by the Swiss National Science Foundation
grant number 200021 181978/1, ``SESAM - Sensing and Sampling: Theory and Algorithms''.
}}
\address{Audiovisual Communications Laboratory\\\'{E}cole Polytechnique F\'{e}d\'{e}rale de Lausanne (EPFL)}
\begin{document}
%
\maketitle
\begin{abstract}
In this paper, we present a modular approach for reconstructing lensless measurements. It consists of three components: a newly-proposed pre-processor, a physics-based camera inverter to undo the multiplexing of lensless imaging, and a post-processor. The pre- and post-processors address noise and artifacts unique to lensless imaging before and after camera inversion respectively. By training the three components end-to-end, we obtain a \SI{1.9}{\decibel} increase in PSNR and a \SI{14}{\percent} relative improvement in a perceptual image metric (LPIPS) with respect to previously proposed physics-based methods. We also demonstrate how the proposed pre-processor provides more robustness to input noise, and how an auxiliary loss can improve interpretability.
\end{abstract}
\section{Introduction}
\label{sec:intro}

Lensless imaging systems are computational cameras that replace the lens with a typically thin phase or amplitude mask at a short distance from the sensor~\cite{Boominathan:22}. This design reduces the constraints imposed by designing traditional lensed systems, allowing for a camera that can be light-weight, compact, and cheap. 
However, removing a traditional optical configuration means that a viewable image is not directly formed on the sensor, and that a computational algorithm is needed to reconstruct an image from the highly multiplexed and defocused measurement. 
When selecting the reconstruction algorithm, there is often a tradeoff between performance and computational complexity. Moreover, the availability of labeled data is a key requirement for approaches that tend towards deep learning. 
Monakhova \etal~\cite{Monakhova:19} studied a range of algorithms -- (1) alternating direction method of multipliers (ADMM)~\cite{ADMM} that incorporates a physical forward model through iterative reconstructions, (2) unrolling a few iterations of ADMM with learned hyperparameters~\cite{lista}, (3) a U-Net~\cite{Unet}, and (4) unrolled ADMM follow by a U-Net. 
The latter three approaches are trained on lensless-lensed image pairs. 
They found that incorporating physical information of the system can reduce the amount of data needed for robust performance, while also allowing for interpretability.
Similarly, Khan \etal~\cite{flatnet} jointly train the point spread function (PSF) used within Wiener filtering (for improved de-multiplexing of the camera's mask) and a neural network-based perceptual enhancement component.

In this paper, we propose a modular reconstruction approach for lensless imaging which consists of (1) a novel pre-processing block to better prepare the content before (2) camera inversion which accounts for the system's response to invert its multiplexing (which can be unrolled ADMM, trainable Wiener filtering, etc), and (3) a post-processing block to handle artifacts from the reconstruction algorithm, color correction, and additional enhancement.
During training, we introduce an auxiliary loss from the camera inversion output to the total loss, to improve the interpretability of the intermediate outputs of our modular reconstruction approach.
The source code for training and applying both the baseline and proposed reconstruction approaches are available on GitHub as part of \textit{LenslessPiCam}~\cite{LenslessPiCam},\footnote{Demo notebook: \url{https://go.epfl.ch/lensless-modular}} which is a complete toolkit for lensless imaging hardware and software.

\section{Lensless Imaging}
\label{sec:related_works}


\subsection{Problem formulation}
\label{ssec:lensless_imaging}

Assuming a desired scene is comprised of point sources that are incoherent with each other, a lensless imaging system can be modeled as a linear matrix-vector multiplication with the system matrix $\bm{H}$:
\begin{align}
    \label{eq:forward_gen}
    \bm{y} = \bm{H}\bm{x} + \bm{n},
\end{align}
where $\bm{y}$ and $\bm{x}$ are the vectorized lensless measurement and scene intensity respectively, and $\bm{n}$ is additive noise.
As obtaining $\bm{H}$ would require an expensive calibration process, the point spread functions (PSF) for each point in the scene are approximated as lateral shifts of the on-axis PSF, \ie linear shift-invariance (LSI) is assumed: 
\begin{equation}
\label{eq:forward}
    \bm{y} = \bm{C}\bm{P}\bm{x} + \bm{n},
\end{equation}
where $\bm{P}$ has a Toeplitz structure with each column being shifted version of the on-axis PSF, and $\bm{C}$ crops the image to the sensor size~\cite{Diffuser3D}.

Under a Gaussian noise assumption, the maximum likelihood estimator of $\bm{x}$ is given by minimizing the mean-squared error (MMSE) between the measurement $\bm{y}$ and $\bm{C}\bm{P} \bm{x}$.
Due to the multiplexing characteristic of most lensless camera PSFs, this minimization problem is ill-posed and regularization is needed.
A typical approach is to use a non-negativity and a sparsity constraint in the total variation space~\cite{Diffuser3D,PhlatCam}, yielding the following optimization problem:
\begin{align}
    \label{eq:opt_reg}
    \hat{\bm{x}} = \arg \min_{\bm{x} \ge 0} \frac{1}{2} ||\bm{y} - \bm{C}\bm{P}\bm{x}||_2^2 + \tau ||\bm{\Psi}\bm{x}||_1,
\end{align}
where $\bm{\Psi}$ computes finite differences along 2D.
\cref{eq:opt_reg} can be solved with an iterative optimization algorithm such as ADMM~\cite{ADMM}.

In the spirit of deep learning, another approach is to collect a sufficiently large dataset of lensless-lensed pairs, and train a neural network, \eg U-Net~\cite{Unet}, to approximate the inverse mapping. 
However, this requires significant training resources and a large dataset to achieve good performance.
Alternatively, there exist a range of method between traditional optimization and deep learning,
that can promote consistency with the measurements (\eg by incorporating a forward model as in \cref{eq:forward}) and can require less data. These methods are described below and also inspire the proposed method presented in~\cref{ssec:proposed-method}.

\subsection{Unrolled reconstruction}
\label{ssec:unrolled}
Unrolled algorithms are a small departure from pure optimization algorithms as ADMM and a step towards data-driven approaches~\cite{lista}. 
In unrolled algorithms, a fixed number of iterations of a convex optimization approach are \emph{unrolled} as layers of a neural network, with each layer $k$ having its own hyperparameters, \eg 4 per ADMM iteration, and hyperparameters are trained end-to-end using backpropagation. Since the number of parameters is small, namely a few dozen, it is possible to train such techniques with a small dataset.

While this approach does not have the same theoretical guarantees as classical convex optimization, they can converge faster and achieve better performance than heuristically selecting hyperparameters. Therefore, the cost of training can be offset by the gain in inference time. In the context of lensless imaging, an unrolled version of ADMM has been explored by Monakhova \etal \cite{Monakhova:19} with great success. They achieve similar results with 5 iterations of unrolled ADMM with learned hyperparameters as with 100 iterations with fixed manually-selected hyperparameters, 
and train with only $100$ examples.
With $23000$ training examples, they also explore adding a learned denoiser (U-Net of 10M parameters at the output of unrolled ADMM), which produces their best results.

\section{Proposed method}
\label{ssec:proposed-method}


\begin{figure}[t!]
    \centering
    \includegraphics[width=\linewidth]{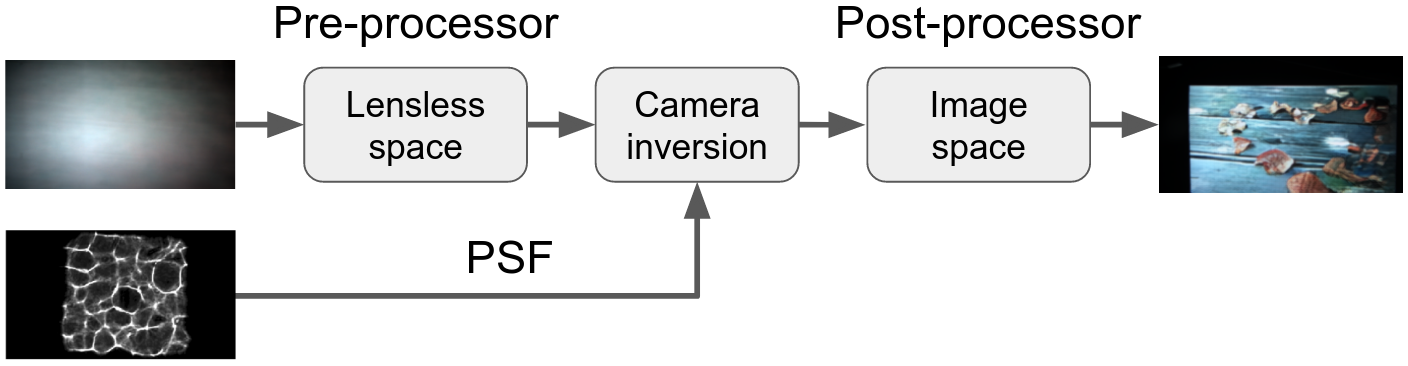}
  \caption{Proposed lensless imaging pipeline.}
  \label{fig:denoise}
\end{figure}

Our proposed imaging pipeline is shown in \cref{fig:denoise}, which has (1) a novel pre-processor to reduce the measurement noise and better prepare the measurement for (2) the subsequent camera inversion and (3) a post-processor to reduce the artifacts from the reconstruction algorithm and perform additional enhancement. 
The pre-processor works in a space with well-studied sensor noise sources, but due the multiplexing property of lensless cameras, the measurements lack any of the structure usually expected by most denoisers. 
On the other hand, the post-processor works in the well-known space of viewable images, but the artifacts from the reconstruction algorithm are complex with very high spatial coherency and cannot be found in typical noisy natural images.
Both processors have a very challenging task and are difficult to design, in particularly simultaneously. 
To this end, our solution is to train both processors \textit{and} the camera inversion end-to-end, such that the appropriate processing can be learned from the data rather than heuristically-designed processing. 
While previous work has demonstrated the effectiveness of using a reconstruction algorithm with a post-processor \cite{Monakhova:19,flatnet}, our experiments reveal the additional benefit of incorporating a pre-processor. 
There are several options for the camera inversion block that can cater to the requirements and constraints of the problem at hand (\eg memory, speed, performance); these approaches include iterative algorithms (ADMM~\cite{ADMM} and the fast iterative shrinkage-thresholding algorithm (FISTA)~\cite{beck2009fast}), unrolling iterative algorithms~\cite{Monakhova:19}, trainable Wiener filtering~\cite{flatnet}, and Tikhonov regularization~\cite{flatcam}.
The following sub-sections further describe the motivation behind the proposed components in our reconstruction approach and during training.

\subsection{Pre- and post-processors}

Lensless imaging can be sensitive to noise and to model mismatch, \eg in the PSF.
In solving~\cref{eq:opt_reg} with ADMM and a noisy PSF estimate $\bm{\hat{P}}$ such that $ \bm{P} = \bm{\hat{P}} + \bm{\Delta}_{\bm{\bm{P}}} $, the authors of~\cite{9546648} demonstrate that \textit{each iteration} update of ADMM is a function of (1) the noisy update that uses $\bm{\hat{P}}$ and (2) error terms from previous iterations.
Inserting the measurement definition $\bm{y}$ from~\cref{eq:forward} into Eq.~15 of~\cite{9546648}, we can observe a perturbation that arises from model mismatch and noise amplification:
\begin{align}
    \bm{x}^{(k)} &= \bm{W}_1 \bm{\hat{x}}^{(k)} + \underbrace{\bm{W}_2 \bm{C}^T \bm{C}\bm{P}\bm{x} +\bm{\epsilon}^{(k-1)}}_{\text{model mismatch}} + \underbrace{\bm{W}_2 \bm{C}^T \bm{n}}_{\text{noise amplification}} \label{eq:noiy_admm_update}
\end{align}
where $\bm{\hat{x}}^{(k)}$ is the noisy ADMM update at the $k^{th}$ iteration, $\bm{\epsilon}^{(k-1)}$ are error terms from the previous iteration, $\{\rho_x, \rho_y, \rho_z\}$ are ADMM hyperparameters, and: 
\begin{align}
\bm{W}_1 &= \left( \bm{W}_3 + \rho_x \delta_{\bm{P}} \right)^{-1} \bm{W}_3,\\
\bm{W}_2 &= (\bm{W}_3 + \rho_x \delta_{\bm{P}})^{-1} \Delta_{\bm{P}}^T \rho_x (\bm{C}^T\bm{C} + \rho_x \bm{I})^{-1} \\
\bm{W}_3 &= \rho_x \bm{\hat{P}}^T \bm{\hat{P}} + \rho_z \bm{C}^T\bm{C} + \rho_y \bm{I} \\
\delta_{\bm{P}} &= \left( \bm{\Delta}_H^T\bm{P} + \bm{\hat{P}}^T \bm{\Delta}_P \right).
\end{align}
From  \cref{eq:noiy_admm_update} we see a clear motivation for the pre- and post-processors:
for the \textit{pre-processor} to (1) minimize the noise $\bm{n}$ that could be amplified at each iteration and (2) to use non-linear operations to project the measurement to a space where the assumptions of LSI better hold for the camera inversion step;
and for the \textit{post-processor} to remove the noise at the output of camera inversion due to model mismatch and noise amplification.
The pre-processor can also improve robustness to variations in input noise.
In \cref{results:pre}, we perform an experiment to show the benefits of incorporating a pre-processor as the input signal-to-noise ratio (SNR) varies.

\subsection{Auxiliary loss}

During training, deep neural networks can experience vanishing/exploding gradients at the intermediate layers. 
An auxiliary loss, \ie having an intermediate output redirected to the final backprogated loss, can lead to more stable training and allow for the training of deeper neural networks~\cite{szegedy2015going}.
For our application, we redirect the output of the camera inversion block to the output loss, as this image could be discernible and a loss can be computed with the ground-truth lensed image. 
The proposed auxiliary loss can also ensure more interpretable results, \eg between the reconstruction algorithm and the post-processor, as shown in the experiment of~\cref{results:res}.

\section{Experimental setup}
\label{sec:experiments}

\subsection{Dataset}
\label{ssec:dataset}
We use the DiffuserCam Lensless Mirflickr (DLM) dataset~\cite{Monakhova:19}, which consists of 25000 pairs of standard (lensed) images and their associated lensless measurement with a diffuser-based lensless camera.
The dataset is collected by projecting an image on a computer monitor, and simultaneously capturing an image with both a lensed and a lensless camera.
We follow the train-test split suggested by the authors: the first 1000 image pairs are used as a test set, and the remaining 24000 pairs are used for training.
In our experiments, we downsample the images by $2\times$ to a resolution of $240 \times135 $ pixels. 

\subsection{Training and evaluation}
\label{ssec:training}
All experiments are run on a Dell Precision 5820 Tower X-Series (08B1) machine with an Intel i9-10900X CPU and an NVIDIA RTX A5000 GPU. PyTorch~\cite{Pytorch} is used for dataset preparation and training.
All models are trained for 25 epochs on the 24000 image pairs with a batch size of 8. The Adam optimizer~\cite{Adam} is used with a learning rate of $10^{-4}$. The loss function is a sum of the MSE  and the LPIPS score (with VGG weights) \cite{LPIPS} between the model output and the lensed image:
\begin{equation}
    \mathscr{L}\left(\bm{x},\bm{\hat{x}}\right) = \mathscr{L}_{\text{MSE}}\left(\bm{x},\bm{\hat{x}}\right) + \mathscr{L}_{\text{LPIPS}}\left(\bm{x},\bm{\hat{x}}\right).
\end{equation}
With the proposed auxiliary loss with weighting $\alpha > 0$, the backpropagated loss (when there is a post-processor) is:
\begin{equation}
    \label{eq:loss_res}
    \mathscr{L}_{\text{res}}\left(\bm{x},\bm{\hat{x}},\bm{\hat{x}}_{\text{inv}}\right) = \mathscr{L}\left(\bm{x},\bm{\hat{x}}\right) + \alpha \hspace{0.2em} \mathscr{L}\left(\bm{x},\bm{\hat{x}}_{\text{inv}}\right).
\end{equation}
For evaluation, we use peak signal-to-noise ratio (PSNR) and LPIPS. The former is measured in decibels (higher is better) while the latter is within $[0,1]$ and lower is better.

\subsection{Reconstruction algorithm and processors}
\label{ssec:model}

\begin{figure}[t!]
    \centering
    \includegraphics[width=0.95\linewidth]{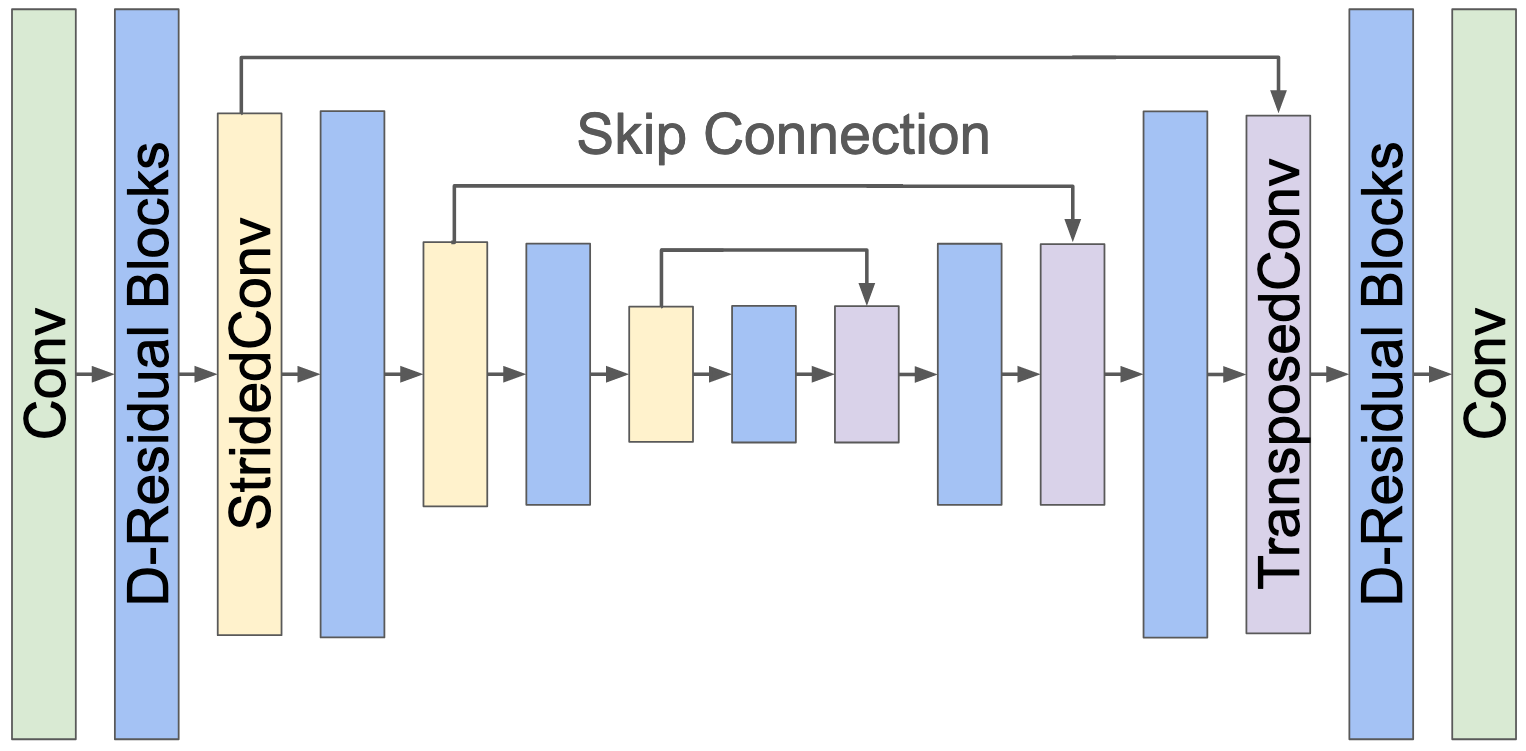}
  \caption{Architecture of pre- and post-processors.}
  \label{fig:unetres}
\end{figure}

Unless otherwise noted, all baseline and proposed approaches use ADMM for camera inversion (see \cref{fig:denoise})  with different variations: 100 iterations with fixed hyperparameters (\textit{ADMM100}), or unrolled with 20 iterations (\textit{Unrolled20}).
The processor components are U-Nets with $d$ residual blocks between each down- and up-sampling layer, as shown in \cref{fig:unetres}. 
There are three down-sampling layers using strided convolutions with a stride of two, and three up-sampling layers using transposed convolutions. The number of channels is 64 for the first layer and is doubled at each down-sampling layer. 
As processors, we either train the above U-Net architecture from scratch with $d=2$, or fine-tune DRUNet~\cite{DruNet} (above architecture with $d=4$ that has been trained on degraded-clean image pairs).
In our results, we refer to these processors as \textit{UNet2} and \textit{DRUNet} respectively.

\section{Results}
\label{ssec:results}

We perform three experiments to demonstrate the benefits of the proposed techniques:
\begin{enumerate}
    \item In \cref{results:pre}, we vary the input SNR to show the utility of the proposed pre-processor component.
    \item In \cref{results:res}, we show how the auxiliary loss from the reconstruction output can improve interpretability.
    \item In \cref{results:ablation}, we perform a comparison with baselines from previous work.
\end{enumerate}

\subsection{Varying input signal-to-noise ratio}
\label{results:pre}

\begin{figure}[t!]
    \centering
	\begin{subfigure}{0.49\linewidth}
		\centering
		\includegraphics[width=0.99\linewidth]{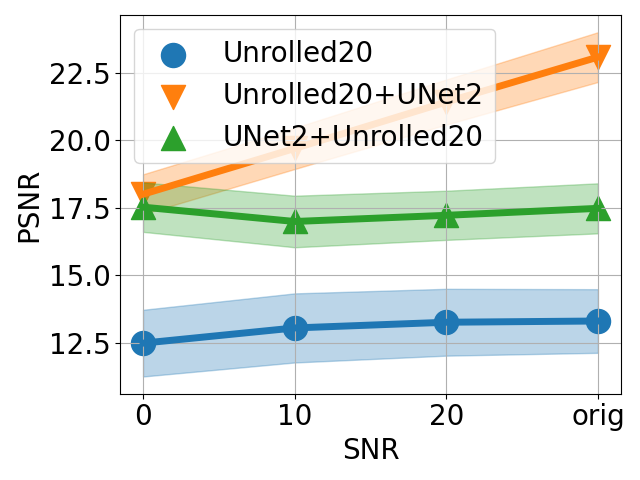} 
		\caption{PSNR.}
		\label{fig:exp1_PSNR}
	\end{subfigure}
	\begin{subfigure}{0.49\linewidth}
		\centering
		\includegraphics[width=0.99\linewidth]{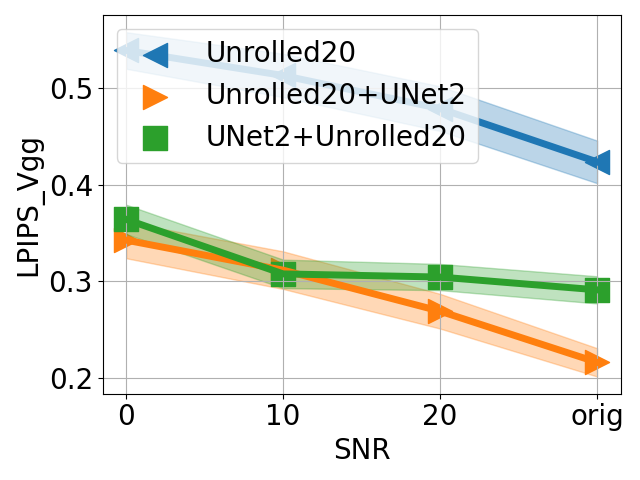}
        \caption{LPIPS.}
		\label{fig:exp1_LPIPS_Vgg}
	\end{subfigure}
	\caption{Average and standard deviation (shaded) performance on test set for varying signal-to-noise ratio; ``orig'' (x-axis) means no simulated noise is added.}
	\label{fig:vary_snr}
\end{figure}

\begin{figure}[t!]
    \centering
	\begin{subfigure}{0.49\linewidth}
		\centering
		\includegraphics[width=0.99\linewidth]{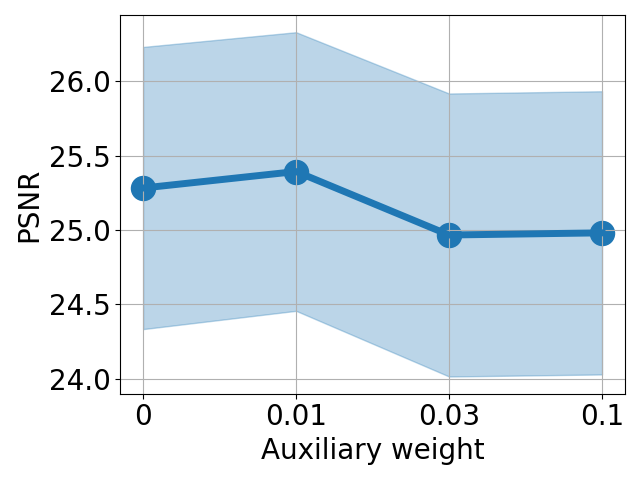} 
		\caption{PSNR.}
		\label{fig:exp2_PSNR}
	\end{subfigure}
	\begin{subfigure}{0.49\linewidth}
		\centering
		\includegraphics[width=0.99\linewidth]{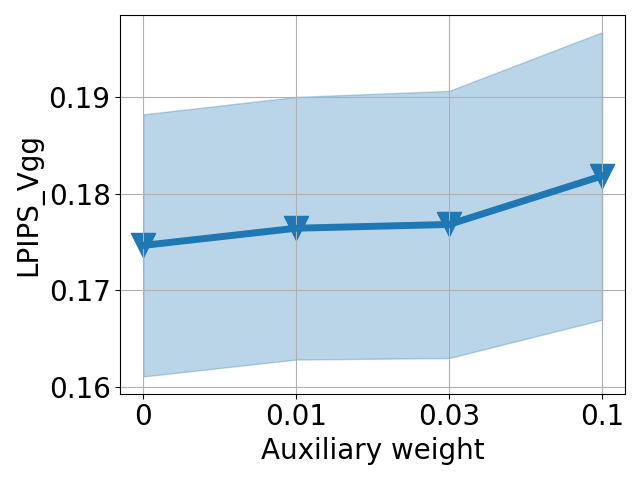}
        \caption{LPIPS.}
		\label{fig:exp2_LPIPS_Vgg}
	\end{subfigure}
	\caption{Average and standard deviation (shaded) performance on test set for varying auxiliary loss weight.}
	\label{fig:vary_alpha}
 \vspace{-0.4cm}
\end{figure}

\newcommand{\figsizebench}{0.16}
\begin{figure*}[t!]
\centering
	\begingroup
	\renewcommand{\arraystretch}{1} 
	\setlength{\tabcolsep}{0.2em} 
	\begin{tabular}{ccccc}
    \makecell{Lensed\\(clean)} & \makecell{Post-proc.\\ (\SI{0}{\decibel})} & \makecell{Pre-proc.\\ (\SI{0}{\decibel})} & \makecell{Post-proc.\\ (\SI{20}{\decibel})} & \makecell{Pre-proc.\\ (\SI{20}{\decibel})}\\
    \insertwithsubimagenew[88 17 100 90]{\figsizebench\linewidth}{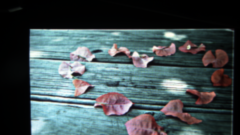} & 
    \insertwithsubimagenew[88 17 100 90]{\figsizebench\linewidth}{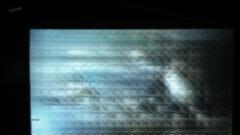} & 
    \insertwithsubimagenew[88 17 100 90]{\figsizebench\linewidth}{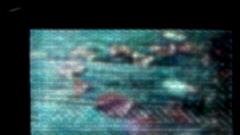} & 
    \insertwithsubimagenew[88 17 100 90]{\figsizebench\linewidth}{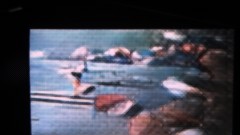} & 
    \insertwithsubimagenew[88 17 100 90]{\figsizebench\linewidth}{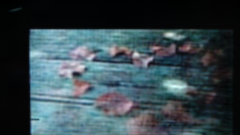} \\
    \insertwithsubimagenew[98 72 90 35]{\figsizebench\linewidth}{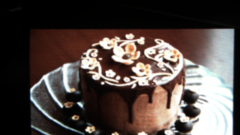} & 
    \insertwithsubimagenew[98 72 90 35]{\figsizebench\linewidth}{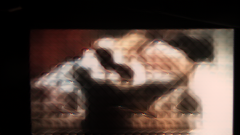} & 
    \insertwithsubimagenew[98 72 90 35]{\figsizebench\linewidth}{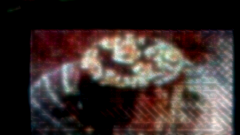} & 
    \insertwithsubimagenew[98 72 90 35]{\figsizebench\linewidth}{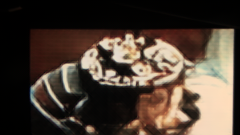} & 
    \insertwithsubimagenew[98 72 90 35]{\figsizebench\linewidth}{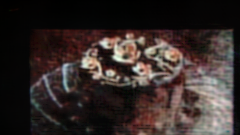}  \\
    \insertwithsubimagenew[88 17 100 90]{\figsizebench\linewidth}{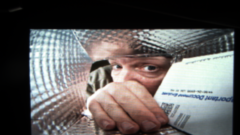} & 
    \insertwithsubimagenew[88 17 100 90]{\figsizebench\linewidth}{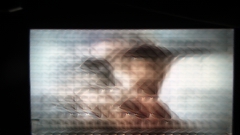} & 
    \insertwithsubimagenew[88 17 100 90]{\figsizebench\linewidth}{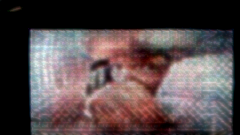} & 
    \insertwithsubimagenew[88 17 100 90]{\figsizebench\linewidth}{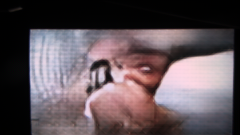} & 
    \insertwithsubimagenew[88 17 100 90]{\figsizebench\linewidth}{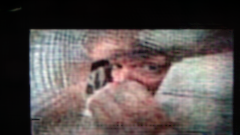}  \\
    \insertwithsubimagenew[73 70 115 37]{\figsizebench\linewidth}{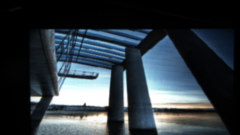} & 
    \insertwithsubimagenew[73 70 115 37]{\figsizebench\linewidth}{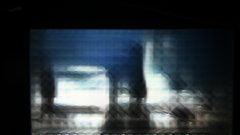} & 
    \insertwithsubimagenew[73 70 115 37]{\figsizebench\linewidth}{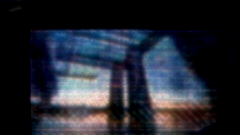} & 
    \insertwithsubimagenew[73 70 115 37]{\figsizebench\linewidth}{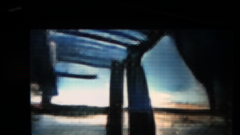} & 
    \insertwithsubimagenew[73 70 115 37]{\figsizebench\linewidth}{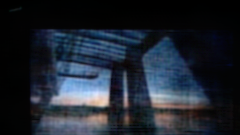}  \\
    \insertwithsubimagenew[34 37 154 70]{\figsizebench\linewidth}{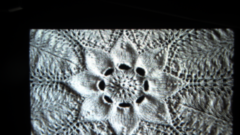} & 
    \insertwithsubimagenew[34 37 154 70]{\figsizebench\linewidth}{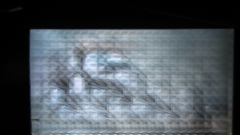} & 
    \insertwithsubimagenew[34 37 154 70]{\figsizebench\linewidth}{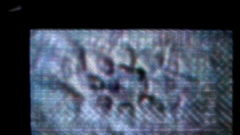} & 
    \insertwithsubimagenew[34 37 154 70]{\figsizebench\linewidth}{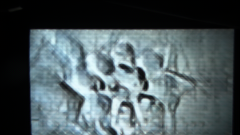} & 
    \insertwithsubimagenew[34 37 154 70]{\figsizebench\linewidth}{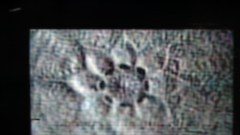}  \\
	\end{tabular}
	\endgroup
	\caption{Example outputs of varying signal-to-noise ratio.}
  \label{fig:exp1_compare}
\end{figure*}

\newcommand{\figsizegen}{0.16}
\newcommand{\newlinegen}{1pt}
\begin{figure*}[t!]
\centering
	\begingroup
	\renewcommand{\arraystretch}{1} 
	\setlength{\tabcolsep}{0.2em} 
	\begin{tabular}{ccccc}
		  \SI{10}{\decibel} & \SI{15}{\decibel} & \textit{\SI{20}{\decibel}} & \SI{25}{\decibel} & \SI{30}{\decibel}\\

    \insertwithsubimagenew[108 57 80 50]{\figsizegen\linewidth}{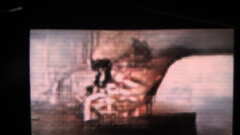}
  & \insertwithsubimagenew[108 57 80 50]{\figsizegen\linewidth}{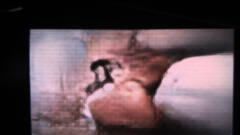}
  &\insertwithsubimagenew[108 57 80 50]{\figsizegen\linewidth}{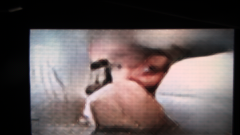}
  & \insertwithsubimagenew[108 57 80 50]{\figsizegen\linewidth}{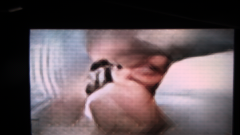}
  & \insertwithsubimagenew[108 57 80 50]{\figsizegen\linewidth}{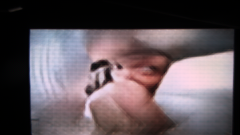}
\\[\newlinegen]
        \insertwithsubimagenew[108 57 80 50]{\figsizegen\linewidth}{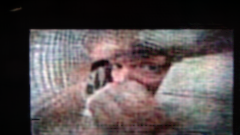}
  & \insertwithsubimagenew[108 57 80 50]{\figsizegen\linewidth}{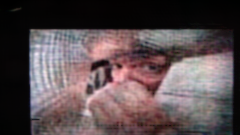}
  & \insertwithsubimagenew[108 57 80 50]{\figsizegen\linewidth}{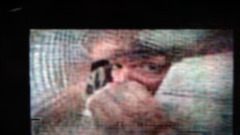}
  & \insertwithsubimagenew[108 57 80 50]{\figsizegen\linewidth}{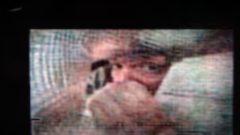}
  & \insertwithsubimagenew[108 57 80 50]{\figsizegen\linewidth}{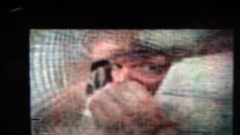}
\\
	\end{tabular}
	\endgroup
	\caption{Applying \textit{Unrolled20+UNet2} (top) and \textit{UNet2+Unrolled20} (bottom), both trained at \SI{20}{\decibel}, to various noise levels.}
  \label{fig:gensnr}
\end{figure*}

With this experiment, we show the effectiveness and necessity of a pre-processor for removing noise prior to the camera inversion.
To this end, we add shot noise (\ie Poisson distribution) to the DLM dataset at various SNRs (\SI{0}{\decibel}, \SI{10}{\decibel}, \SI{20}{\decibel}), and train the following models on the noisy DLM dataset: \textit{Unrolled20}, \textit{Unrolled20+UNet2} (just post-processor as in~\cite{Monakhova:19}), \textit{UNet2+Unrolled20} (just pre-processor).

\cref{fig:vary_snr} shows the PSNR and LPIPS of the three approaches as the input SNR varies.
Using a pre-processor (\textit{UNet2+Unrolled20}) is more robust to variations in the input SNR, as its slope in \cref{fig:vary_snr} is less steep.
While the metrics for using a post-processor (\textit{Unrolled20+UNet2}) are better, the reconstructions, as shown in~\cref{fig:exp1_compare}, reveal the limitations of relying on these metrics for evaluating performance.
We observe that using a pre-processor (third and last column) is better at recovering finer details than the corresponding outputs when using a post-processor (second and fourth column).
The lower image quality metrics for the pre-processor are likely due to the grainier outputs and/or color differences with respect to the \textit{Lensed} image.
However, these can be easily remedied with the addition of a post-processor.
In other words, this experiment reveals the complementary nature of the pre- and post-processors: the pre-processor allows finer details to be recovered by the camera inversion, while the latter enhances the global quality of the image at the output of the camera inversion.
We also find that using the pre-processor is more robust to differences between the SNR at training and at inference, \ie generalizing to unseen noise levels. 
As seen in \cref{fig:gensnr}, for both lower and higher SNRs (than used at training) the reconstruction is more robust with the pre-processor.
In the following experiments, no noise is added to the original DLM dataset.

\newcommand{\figsizeaux}{0.17}
\newcommand{\newlineaux}{18pt}
\begin{figure*}[t!]
\centering
	\begingroup
	\renewcommand{\arraystretch}{1} 
	\setlength{\tabcolsep}{0.1em} 
	\begin{tabular}{cccc|c}
		  & After pre-processor & After camera inversion & After post-processor & Lensless/ Lensed \\

    $\alpha=0$ \hspace{0.5em} & \includegraphics[width=\figsizeaux\linewidth,valign=m]{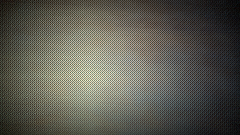}
  & \includegraphics[width=\figsizeaux\linewidth,valign=m]{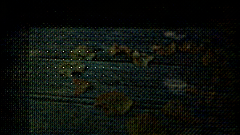}
  &\includegraphics[width=\figsizeaux\linewidth,valign=m]{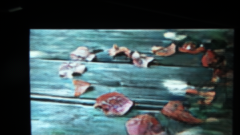}
  & \includegraphics[width=\figsizeaux\linewidth,valign=m]{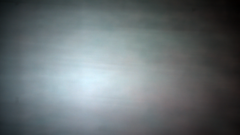}
\\[\newlineaux]
  
		$\alpha=0.1$ \hspace{0.5em} & \includegraphics[width=\figsizeaux\linewidth,valign=m]{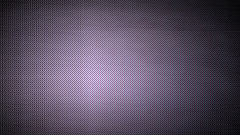}
  & \includegraphics[width=\figsizeaux\linewidth,valign=m]{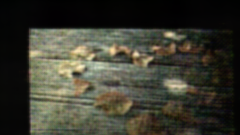}
  & \includegraphics[width=\figsizeaux\linewidth,valign=m]{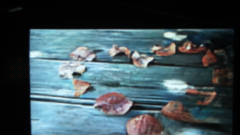}
  & \includegraphics[width=\figsizeaux\linewidth,valign=m]{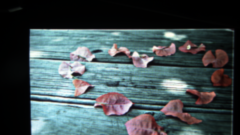}
\\
	\end{tabular}
	\endgroup
	\caption{Influence of auxiliary loss from reconstruction output: top row does not use the auxiliary loss during training; bottom row uses the auxiliary loss with a weight of $\alpha=0.1$.}
  \label{fig:residual}
\end{figure*}

\subsection{Auxiliary loss}
\label{results:res}

With this experiment, we evaluate the effect of incorporating an auxiliary loss of the unrolled output to the total loss.
To this end, we add different amounts of the unrolled output loss, \ie $\alpha = [0, 0.01, 0.03, 0.1]$ for \cref{eq:loss_res} ($\alpha=0$ is equivalent to no auxiliary loss).
The goal is to improve the interpretability of the reconstruction approach, such that insight can be gained from the intermediate outputs.
\cref{fig:vary_alpha} shows the PSNR and LPIPS of a combined pre- and post-processor approach (\textit{UNet2+Unrolled20+UNet2}) as $\alpha$ varies.
The metrics slightly deteriorate as $\alpha$ increases, which is expected 
as the image prior to the post-processor is poorer in quality.
Nonetheless, this decrease in image quality metrics is minimal: \SI{0.301}{\decibel} in PSNR and \SI{4.13}{\percent} relative increase in LPIPS from $\alpha=0$ to $\alpha=0.1$.
\cref{fig:residual} shows an example of intermediate outputs with and without the auxiliary loss during training.
The final outputs appear very similar with and without the auxiliary loss, whereas the image after the camera inversion is more discernible with the auxiliary loss.
From these intermediate outputs, we can observe similar results as before: the pre-processor prepares the input image for improved camera inversion while the post-processor puts the finishing touches.

\subsection{Benchmark with previous work}
\label{results:ablation}

\begin{figure*}[t!]
\centering
	\begingroup
	\renewcommand{\arraystretch}{1} 
	\setlength{\tabcolsep}{0.2em} 
	\begin{tabular}{cccccc}
    Plug-and-play~\cite{pnp} & 
    Unrolled20~\cite{Monakhova:19} &\makecell{Unrolled20\\+DRUNet~\cite{Monakhova:19}}& \makecell{TrainInv\\+DRUNet~\cite{flatnet}} & \makecell{UNet2+Unrolled20\\+UNet2 (Proposed)} & Lensed \\
    
    \insertwithsubimagenew[88 17 100 90]{\figsizebench\linewidth}{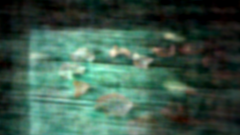} & 
    \insertwithsubimagenew[88 17 100 90]{\figsizebench\linewidth}{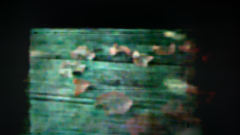} & 
    \insertwithsubimagenew[88 17 100 90]{\figsizebench\linewidth}{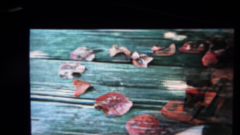} & 
    \insertwithsubimagenew[88 17 100 90]{\figsizebench\linewidth}{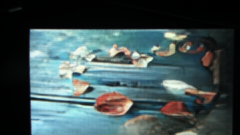} & 
    \insertwithsubimagenew[88 17 100 90]{\figsizebench\linewidth}{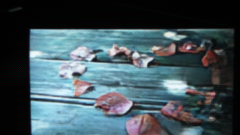} & 
    \insertwithsubimagenew[88 17 100 90]{\figsizebench\linewidth}{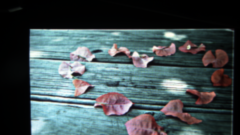} \\
    
    \insertwithsubimagenew[98 72 90 35]{\figsizebench\linewidth}{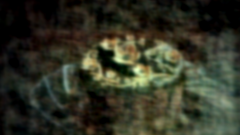} & 
    \insertwithsubimagenew[98 72 90 35]{\figsizebench\linewidth}{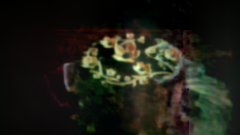} & 
    \insertwithsubimagenew[98 72 90 35]{\figsizebench\linewidth}{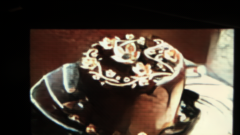} & 
    \insertwithsubimagenew[98 72 90 35]{\figsizebench\linewidth}{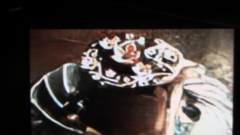} & 
    \insertwithsubimagenew[98 72 90 35]{\figsizebench\linewidth}{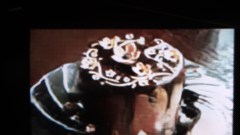} & 
    \insertwithsubimagenew[98 72 90 35]{\figsizebench\linewidth}{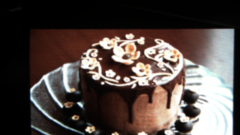} \\
    
    \insertwithsubimagenew[88 17 100 90]{\figsizebench\linewidth}{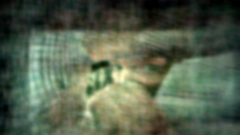} & 
    \insertwithsubimagenew[88 17 100 90]{\figsizebench\linewidth}{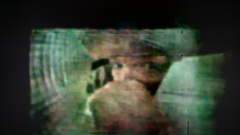} & 
    \insertwithsubimagenew[88 17 100 90]{\figsizebench\linewidth}{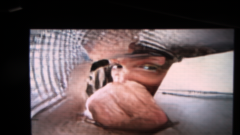} & 
    \insertwithsubimagenew[88 17 100 90]{\figsizebench\linewidth}{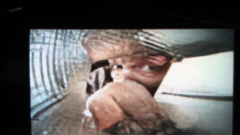} & 
    \insertwithsubimagenew[88 17 100 90]{\figsizebench\linewidth}{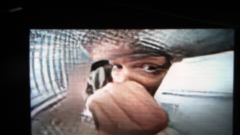} & 
    \insertwithsubimagenew[88 17 100 90]{\figsizebench\linewidth}{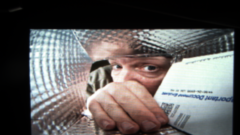} \\
    
    \insertwithsubimagenew[73 70 115 37]{\figsizebench\linewidth}{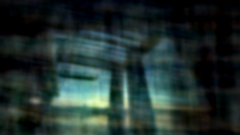} & 
    \insertwithsubimagenew[73 70 115 37]{\figsizebench\linewidth}{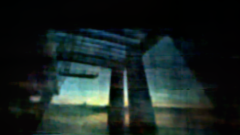} & 
    \insertwithsubimagenew[73 70 115 37]{\figsizebench\linewidth}{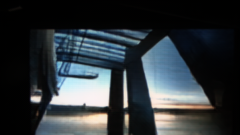} & 
    \insertwithsubimagenew[73 70 115 37]{\figsizebench\linewidth}{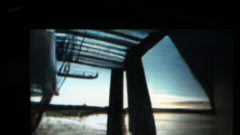} & 
    \insertwithsubimagenew[73 70 115 37]{\figsizebench\linewidth}{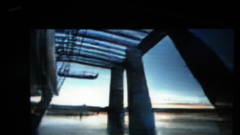} & 
    \insertwithsubimagenew[73 70 115 37]{\figsizebench\linewidth}{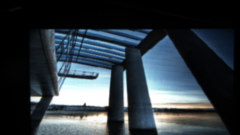} \\
    
    \insertwithsubimagenew[34 37 154 70]{\figsizebench\linewidth}{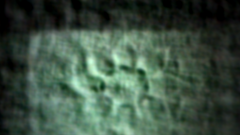} & 
    \insertwithsubimagenew[34 37 154 70]{\figsizebench\linewidth}{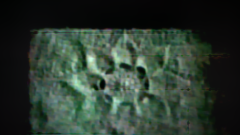} & 
    \insertwithsubimagenew[34 37 154 70]{\figsizebench\linewidth}{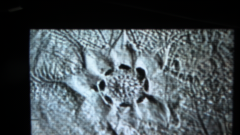} & 
    \insertwithsubimagenew[34 37 154 70]{\figsizebench\linewidth}{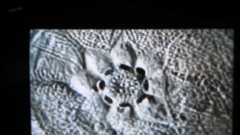} & 
    \insertwithsubimagenew[34 37 154 70]{\figsizebench\linewidth}{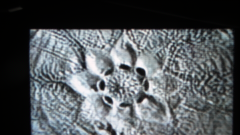} & 
    \insertwithsubimagenew[34 37 154 70]{\figsizebench\linewidth}{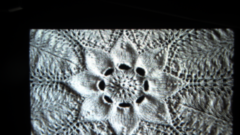} \\
	\end{tabular}
	\endgroup
	\caption{Results of different reconstruction approaches.}
  \label{fig:benchmark}
\end{figure*}

\begin{table}[t!]
  \centering
  \scalebox{0.82}{
  \begin{tabular}{|c|c|c|c|c|c|}
    \hline
    Method & PSNR $\uparrow$           & LPIPS  $\downarrow$   & \makecell{\# learnable \\paramaters} & \makecell{Inference\\time [ms]} \\
    \hline
    ADMM100~\cite{Diffuser3D}                      &     15.2   &   0.547    &  - & 56.0\\\hline
    Plug-and-play~\cite{pnp} & 14.9 & 0.590 & - &  549 \\\hline
    Unrolled20~\cite{Monakhova:19}   &  13.3         &    0.424    &  80  & 13.4 \\\hline
    \makecell{Unrolled20\\+DRUNet~\cite{Monakhova:19} } 
     &     23.4    &      0.204    &  32.6M   & 22.4 \\
     \hline
    TrainInv+DRUNet~\cite{flatnet}  &    21.7    &   0.246      &   32.7M  & 8.99 \\
    \hline \hline
    \makecell{UNet2+Unrolled20\\+UNet2 }
     &     \textbf{25.3}    &   \textbf{0.175 }        &  34.0M  &   23.1\\\hline
    \makecell{UNet2+TrainInv\\+UNet2 }
     &    22.7   &    0.224   &  34.1M  &  9.75\\\hline
  \end{tabular}}
  \caption{Results of different models on the test set.}
  \label{tab:ablation}
  \vspace{-0.5cm}
\end{table}

In this final experiment, we compare the proposed reconstruction pipeline with baselines from previous work.
\cref{tab:ablation} presents our results, and \cref{fig:benchmark} compares reconstructions on test set images. The best result is obtained with the proposed pre- and post-processing approach (last rows in \cref{tab:ablation} and last column in \cref{fig:benchmark}), with \textit{UNet2} as a pre-processor and \textit{UNet2} as the post-processor.

While the first two approaches in \cref{tab:ablation} require no training data, they miss the necessary enhancement to address artifacts of lensless imaging and reconstruction. 
\textit{Plug-and-play} applies 20 iterations of ADMM and replaces the soft-thresholding proximal step with \textit{DRUNet} (without having to fine-tune), but \textit{DRUNet} has only been trained with Gaussian noise and cannot deal with the artifacts either.
With training data, \textit{Unrolled20} significantly improves LPIPS with respect to \textit{ADMM100} (PSNR is worse as MSE and LPIPS are equally weighted in the loss); but it cannot address color correction.
\textit{Unrolled20+DRUNet} is similar to \textit{Le-ADMM-U} from~\cite{Monakhova:19}; the added processor (fine-tuned \textit{DRUNet}) significantly improves performance.
\textit{TrainInv+DRUNet}, where \textit{TrainInv} fine-tunes the PSF used for inversion as in~\cite{flatnet}, also exhibits improved performance compared to \textit{Unrolled20}. 
However, \textit{TrainInv} is less performant as a camera inverter than \textit{Unrolled20} as the latter uses multiple iterations, but it also makes \textit{TrainInv} faster (more than $2\times$).
For both  camera inversion approaches, we obtain the best results by splitting the enhancement parameters between the pre- and post-processors as proposed in this work (last rows of \cref{tab:ablation}).

\section{Conclusion}
\label{sec:conclusion}

In this paper, we propose a novel lensless image reconstruction pipeline. It consists of three components trained end-to-end: (1) a newly-proposed pre-processor, (2) a camera inversion block, \eg unrolled ADMM to incorporate knowledge of the physical system, and (3) a post-processor. 
Our results demonstrate the robustness of the system to a wide range of SNRs, and the effectiveness of the proposed pipeline: \SI{1.9}{\decibel} increase in PSNR and \SI{14}{\percent} relative improvement in LPIPS with respect to previously proposed physics-based methods.
Reproducibility and well-structured code are also a priority: the baseline and proposed techniques have been incorporated into \textit{LenslessPiCam}~\cite{LenslessPiCam} in an object-oriented fashion.

For future work, it would be interesting to jointly optimize the PSF and reconstruction parameters for a specific task~\cite{khan23}.
Moreover, we saw the generalizability of the pre-processor to different SNRs than at training. 
It would be interesting to explore how well the proposed modular pipeline generalizes for other lensless systems, and if the components trained for one system can generalize to other systems (without re-training).

\pagebreak
\bibliographystyle{IEEEbib}
\bibliography{refs}

\end{document}